\documentclass[preprint]{revtex4}

\usepackage{psfig}
\usepackage{epsfig}
\usepackage{graphicx}
\usepackage{dcolumn}
\usepackage{amsmath}

\makeatletter
\def\btt#1{\texttt{\@backslashchar#1}}%
\DeclareRobustCommand\bblash{\btt{\@backslashchar}}%
\makeatother

\begin{document}

\title{Tachyon Vortex *}

\author{Dao-jun Liu}
\author{Xin-zhou Li}\email{kychz@shtu.edu.cn}
\thanks{\\\hspace{1.5cm} *Supported by National Nature Science Foundation of China
under Grant No. 19875016 and Foundation of Shanghai Development
for Science and Technology under Grant No. 01JC14035.}

\affiliation{ Shanghai United Center for Astrophysics, Shanghai
Normal University,
Shanghai 200234 ,China}%

\date{\today}

\begin{abstract}
The property and gravitational field of global string of tachyon
matter are investigated in a four dimentional approximately
cylindrically-symmetric spacetime with a deficit angle.
Especially, we give an exact solution of the tachyon field in the
flat spacetime background and we also find the solution of the
metric in the linearized approximation of gravity.
\end{abstract}


\maketitle

Various topological defects such as domain wall, string(vortex)
and monopole could be formed by the symmetry-breaking  phase
transitions in the early universe and their existence has
important implications in cosmology\cite{Kibble, vilenkin}. The
symmetry breaking model of ordinary scalar field\cite{Chen} can be
prototypically written as

\begin{equation}
\emph{L}
=\frac{1}{2}\partial_{\mu}\phi^{a}\partial^{\mu}\phi^{a}-V(f)
\end{equation}

\noindent where $\phi^{a}$ is a set of scalar fields, $a=1, ...,
N$, $f=(\phi^{a}\phi^{a})^{\frac{1}{2}}$. The model has O(N)
symmetry and admits domain wall, string and monopole solutions for
$N=1, 2$ and $3$, respectively.  Usually, the potential $V(f)$ has
a minimum at a finite non-zero value of $f$. On the other hand, in
Ref.\cite{Cho1, Cho2}, Cho and Vilenkin investigated the defects
in models where $V(f)$ has a local maximum at $f=0$ but no minima;
instead, it monotonically decrease to zero at $f\rightarrow
\infty$. And they called this kind of defects "vacuumless
defects". The main gravitational property of all the defects
mentioned above is the divergent mass (or mass density), which
leads to an effect of deficit angle and negative mass (or mass
density)\cite{Shi, Li1}.

Recently, pioneered by Sen \cite{Sen1}, the study of non-BPS
objects such as non-BPS branes, brane-antibrane configurations or
space-like branes \cite{Strominger} has been attracting physical
interests in string theory. Sen showed that classical decay of
unstable D-brane in string theories produces pressureless gas with
non-zero energy density. The basic idea is that the usual open
string vacuum is unstable but there exists a stable vacuum with
zero energy density. There is evidence that this state is
associated with the condensation of electric flux tubes of closed
string \cite{Sen2}. These flux tubes described successfully using
an effective Born-Infeld action \cite{Sen3}.  The tachyon rolling
towards its minimum at infinity as a dark matter candidate was
also proposed by Sen\cite{Sen2}. Gibbons took into account
coupling to gravitational field by adding an Einstein-Hilbert term
to the effective action of the tachyon on a brane, and  initiated
a study of "tachyon cosmology" \cite{Gibbons}. Several authors
have investigated the process of rolling of the tachyon in the
cosmological background\cite{Li4, Li5}.

Especially, in the case of merging $D$-$\bar{D}$ branes, the
tachyon field is complex and the potential has a phase symmetry
$V(T)=V(Te^{i\alpha})$, the cosmic strings would be produced after
the annihilation of the branes \cite{Sen}.

It is therefore of importance to investigate the property and the
gravity of the topological defects of tachyon field in the
stationary spacetime background. In this paper, we first present
the rolling tachyon dynamics in the stationary spacetime, and then
discuss the property of global string of the tachyon field in the
flat spacetime and in the curved spacetime respectively.

A general static, cylindrically-symmetric metric can be
represented as

\begin{equation}\label{metric}
ds^2=B(r)(dt^2-dz^2)-dr^2-r^2A(r)d\theta^2.
\end{equation}

\noindent And the Lagrangian density of rolling tachyon with
potential $V(\phi)$, which couples to the Einstein gravity, can be
written as the following Born-Infeld form:

\begin{eqnarray}\label{LD:2}
L&=&L_R+L_T\nonumber\\
&=&\sqrt{-g}\bigg[\frac{R}{2\kappa}-V(|T|)\sqrt{1-g^{\mu\nu}\partial_{\mu}T\partial_{\nu}T^{\dagger}}\bigg]
\end{eqnarray}

\noindent where $T$ is  a complex tachyon field, and $g_{\mu\nu}$
is the metric. The field configuration describing a string(vortex)
is given by

\begin{equation}\label{ansatz:3}
T=f(r)e^{i\theta}, \hspace{0.8cm} T^{\dagger}=f(r)e^{-i\theta}.
\end{equation}

\noindent Using Eqs.(\ref{metric})-(\ref{ansatz:3}), we can obtain
the following Euler-Lagrange equation:

\begin{eqnarray}\label{ELequation1}
\frac{1}{V}\frac{dV}{df}+\frac{f}{Ar^2}&=&f''+(\frac{B'}{B}+\frac{A'}{2A}+\frac{1}{r})f'\nonumber\\
&-&f'\frac{f'f''+\frac{f'f}{Ar^2}
-\frac{f^2A'}{2Ar^2}-\frac{f^2}{Ar^3}}{1+f'^2+\frac{f^2}{Ar^2}}
\end{eqnarray}

\noindent where the prime denotes the derivative with respect to
$r$ and the Einstein equations reads:

\begin{equation}\label{EinsteinEq1}
-\frac{1B''}{2B}-\frac{B'}{2rB}-\frac{B'A'}{4BA}-\frac{A'}{rA}+\frac{B'^2}{4B^2}-\frac{A''}{2A}=\kappa
T^0_0
\end{equation}

\begin{equation}\label{EinsteinEq2}
-\frac{B'}{rB}-\frac{A'B'}{2AB}-\frac{B'^2}{4B^2}=\kappa T^1_1
\end{equation}

\begin{equation}\label{EinsteinEq3}
\frac{B'^2}{4B^2}-\frac{B''}{B}=\kappa T^2_2
\end{equation}

\noindent where the energy-momentum tensor $T^{\mu}_{\nu}$ of the
system are given by

\begin{equation}
T^0_0=T^3_3=V(f)\sqrt{1+f'^2+\frac{f^2}{Ar^2}}
\end{equation}

\begin{equation}
T^1_1=\frac{V(f)(1+\frac{f^2}{Ar^2})}{\sqrt{1+f'^2+\frac{f^2}{Ar^2}}}
\end{equation}

\begin{equation}
T^2_2=\frac{V(f)(1+f'^2)}{\sqrt{1+f'^2+\frac{f^2}{Ar^2}}}
\end{equation}

\noindent and the rest is zero. Obviously, the configuration of
the system depends on the tachyon potential $V(T)$. According to
Sen \cite{Sen2}, the potential should have an unstable maximum at
$T=0$ and decay exponentially to zero when $T\rightarrow\infty$.
There are lots of functional forms that satisfy the above two
requirements. In the following sections, we choose the tachyon
potential as follows:

\begin{equation}\label{V1}
V(f)=M^4(1+4\lambda f^4)^{1/4}\exp(-\lambda f^4)
\end{equation}

\noindent where $M$ and $\lambda$ are two constants and both are
greater than zero. It is not difficult to find that the potential
 satisfies  the two requirement proposed by Sen.

In the flat spacetime, the Euler-Lagrange equation
(\ref{ELequation1}) can be reduced to the following equation:

\begin{equation}\label{ELequation2}
\frac{1}{V}\frac{dV}{df}+\frac{f}{r^2}=f''+\frac{f'}{r}-f'\frac{f'f''+\frac{f}{r^2}(f'-\frac{f}{r})}{1+f'^2+\frac{f^2}{r^2}},
\end{equation}

\noindent and the energy density of the system $\varepsilon$ is
given by

\begin{equation}
T^0_0=V(f)\sqrt{1+f'^2+\frac{f^2}{r^2}}.
\end{equation}

For the potential (\ref{V1}), the equation (\ref{ELequation2}) has
a simple exact solution

\begin{equation}\label{f1}
f(r)=\lambda^{-\frac{1}{4}}\bigg(\frac{r}{\delta}\bigg)^{-1},
\end{equation}

\noindent where $\delta = (4\lambda)^{-1/4}$ is the size of the
string core, and the corresponding energy density can be written
as

\begin{equation}
T^0_0=M^4\bigg[1+4\bigg(\frac{\delta}{r}\bigg)^4\bigg]^{3/4}\exp\bigg[-\bigg(\frac{\delta}{r}\bigg)^4\bigg].
\end{equation}

The energy per unit length of string at $r\gg \delta$ is

\begin{equation}\label{mu17}
\mu(R)= 2\pi \int^{R}_0 T^0_0 rdr \sim \pi
M^4(R^2-\frac{1}{2\lambda R^2})
\end{equation}

\noindent where the cutoff radius $R$ has the meaning of a
distance to the nearest string (or the loop radius in the case of
a closed loop). We find easily from Eq.(\ref{mu17}) that tachyon
strings are very diffuse objects with most of the energy
distributed at large distances from the string core, and we expect
their spacetime to be substantially distinct from the ordinary
case. They are much more diffuse than ordinary global strings
which have $\mu(R)\propto \ln R$, so that most of the energy is
concentrated near the core. Furthermore, they are similar to the
vacuumless strings which have $\mu(R)\sim (R/\delta)^{4/(n+2)}$,
so that most of the energy is also diffused at large distance
\cite{Cho1}. By analogy with vacuumless strings and ordinary
strings, one can anticipate that tachyon strings will eventually
reach a scaling regime in which the typical distance between the
strings is comparable to the horizon, $R\sim t$ \cite{Cho1}. Let
us now briefly discuss the cosmological evolution of tachyon
strings. The mass per unit length of string is given by
Eq.(\ref{mu17}) where $t$ is substituted for $R$. The relative
contribution of strings to the energy density of the universe is
given by

\begin{equation}
\rho_s/\rho \sim \mu(t)/M^2_p \sim
\pi(\frac{M^2}{M_p})^2(t^2-\frac{1}{2\lambda t^2})
\end{equation}

\noindent where $M_p$ is the Planck mass. It is easy to find the
fraction of energy in strings monotonically grows with time, and
the universe becomes dominated by the strings when $f\sim
\frac{\pi M^4}{2\lambda M^2_p}$. The observed isotropy of the
cosmic microwave background implies $\mu(t_0)/M^2_p \lesssim
10^{-5}$, where $\mu(t_0)$ is the present value of energy per unit
length. Therefore, we have a corresponding constraint on $M$ is $M
\lesssim 1MeV$ as the explanation of structure formation. The
characteristic scale of the observed large scale structure crossed
the horizon at $t\sim t_{eq}\sim 10^{-6}t_0$. The density
fluctuations due to tachyon strings on that scale are of the order
$ \delta \rho_s/\rho_s \sim \mu(t_{eq})/M^2_p \lesssim 10^{-17}$.

In this paper, the spacetime of tachyon strings will be
investigated using the linearized gravity approximation. We shall
first consider the Newtonian approximation. The Newtonian
potential $\Phi$ can be found for the equation

\begin{equation}\label{NewtonEq19}
\nabla^2 \Phi=\frac{\kappa}{2}(T^0_0-T^i_i).
\end{equation}

\noindent For the tachyon string, $f(r)$ is given by Eq.(\ref{f1})
and

\begin{equation}
T^0_0-T^i_i \simeq 2M^4
\end{equation}

\noindent at $r>>\delta$. The solution of Eq. (\ref{NewtonEq19})
is then

\begin{equation}
\Phi(r) \simeq -\frac{M^4}{8\lambda M_p^2f^2}.
\end{equation}

\noindent The linearized approximation applies as long as
$|\Phi(r)|<<1$, which is equivalent to
$f>>\frac{M^2}{\sqrt{8\lambda}M_p}$. Therefore, we should take the
parameters $\lambda$ and $M$ satisfy

\begin{equation}
(4\lambda)^{-1/4}<<R<<\frac{2M_p}{M^2}
\end{equation}

\noindent where $R$ is the distance of string separation (or the
loop radius in the case of closed loop). Next, we express the
metric functions $A(r)$ and $B(r)$ as

\begin{equation}
A(r)=1+\alpha(r), \hspace{1cm} B(r)=1+\beta(r).
\end{equation}

Linearizing in $\alpha(r)$ and $\beta(r)$, and using the flat
space expression (\ref{f1}) for $f(r)$, Eqs.
(\ref{EinsteinEq1})-(\ref{EinsteinEq3}) can be written as follows

\begin{equation}\label{LEE1}
\alpha''+\frac{2\alpha'}{r}=-\frac{\kappa
M^4}{4(r/\delta)^4}\bigg[1+4\bigg(\frac{\delta}{r}\bigg)^4\bigg]^{-1/4}\exp\bigg[-\bigg(\frac{\delta}{r}\bigg)^4\bigg],
\end{equation}

\noindent and

\begin{equation}\label{LEE2}
\beta''+\frac{\beta'}{r}=-2\kappa M^4
\bigg[1+2\bigg(\frac{\delta}{r}\bigg)^4\bigg]\bigg[1+4\bigg(\frac{\delta}{r}\bigg)^4\bigg]^{-1/4}\exp\bigg[-\bigg(\frac{\delta}{r}\bigg)^4\bigg].
\end{equation}

\noindent The solution of external metric is easily found

\begin{equation}\label{metricS26}
ds^2=(1-\frac{\kappa
M^4}{2}r^2)(dt^2-dz^2)-dr^2-r^2(1-\frac{\kappa M^4}{\lambda
r^2})d\theta^2.
\end{equation}

\noindent The metric (\ref{metricS26}) can be expressed by the
form of Newtonian potential

\begin{equation}
ds^2=(1+2\Phi)(dt^2-dz^2)-dr^2-r^2\bigg[1+\bigg(\frac{2\delta}{r}\bigg)^4\Phi\bigg]d\theta^2.
\end{equation}

\noindent Here, we note two qualitative differences between the
metrics of tachyon and ordinary gauge string: (i) for a tachyon
string, the gravitational field is strongly repulsive and the
spacetime becomes singular at a finite distance from the string
core; (ii) the effective deficit angle for a tachyon string

\begin{equation}
\Delta(r)\simeq \pi\kappa M^4(r^2-\frac{1}{2\lambda
r^2})+\frac{\pi}{8}\kappa^2M^8r^2(r^2-\frac{\kappa M^4}{\lambda}),
\end{equation}

\noindent increases with distance from the core, while the
ordinary gauge string deficit angle remains constant.

In this paper, we have studied the property and gravitational
field of global string of tachyon matter(tachyon vortex) in a four
dimensional approximately cylindrically-symmetric spacetime. We
give an exact solution of the tachyon field in the flat spacetime
background and in particular, using the linearized approximation
of gravity, we find a solution of the metric, which denotes a
spacetime with an increasing deficit angle. Contrast to the
ordinary string, the tachyon strings are very diffuse objects with
most of the energy distributed at large distances from the string
core, and their spacetime is substantially distinct from that of
the ordinary string. In this respect, they are more similar to the
vacuumless strings\cite{Cho1, Cho2}. In the $D-\bar{D}$ branes
theory, tachyon field is complex and the potential has a phase
symmetry $V(T) =V(e^{i\alpha}T)$\cite{Sen}. In this model one
would expect formation of the cosmic strings after annihilation of
brane. It is worth noting that although we used the potential
(\ref{V1}) in our discussion, all the conclusions will be
preserved qualitatively if we use the potentials which are
discussed by Sen\cite{Sen2}.

\end{document}